\documentclass[11pt]{elsart}
\usepackage{amsfonts}
\usepackage{amssymb,amsmath} 
\usepackage{mathrsfs} 
\usepackage{graphicx}

\begin{document}
\begin{frontmatter}
\title{The revelation principle fails when the format of each agent's strategy is an action}
\author{Haoyang Wu}
\ead{18621753457@163.com}
\address{Wan-Dou-Miao Research Lab, Shanghai, China.}

\begin{abstract}
In mechanism design theory, a designer would like to implement a social choice function which specifies her favorite outcome for each possible profile of agents' private types. The revelation principle asserts that if a social choice function can be implemented by a mechanism in equilibrium, then there exists a direct mechanism that can truthfully implement it.

This paper aims to propose a failure of the revelation principle. We point out that in any game the format of each agent's strategy is either an informational message or a realistic action, and the action format is very common in many practical cases. The main result is that: For any given social choice function, if the mechanism which implements it has action-format strategies, then ``\emph{honest and obedient}'' will no longer be the Bayesian Nash equilibrium of the direct mechanism, actually the social choice function can only be implemented ``\emph{dishonestly and disobediently}'' in Bayesian Nash equilibrium by the direct mechanism. Consequently, the revelation principle fails when the format of each agent's strategy is an action.
\end{abstract}

\begin{keyword}
Mechanism design; Revelation principle.
\end{keyword}
\end{frontmatter}

\section{Introduction}
In the framework of mechanism design theory \cite{MWG1995, Myerson1982, Narahari2009}, there are one designer and some agents labeled as $1, \cdots, I$. \footnote{In this paper, the designer is always denoted as ``She'', and the agent is denoted as ``He''.} Suppose that the designer would like to implement a social choice function which specifies her favorite outcome for each possible profile of agents' types, and each agent's type is modeled as his privacy. In order to implement a social choice function in equilibrium, the designer constructs a mechanism which specifies each agent's feasible strategy set (\emph{i.e.}, the allowed actions of each agent) and an outcome function ($i.e.$, a rule for how agents' actions get turned into a social choice).

The revelation principle is an important theorem in mechanism design theory. It asserts that if a social choice function can be implemented by a mechanism in equilibrium, then it is truthfully implementable. So far, there have been several criticisms on the revelation principle: Bester and Strausz \cite{Bester2001} pointed out that the revelation principle may fail because of imperfect commitment; Epstein and Peters \cite{Epstein1999} proposed that the revelation principle fails in situations where several mechanism designers compete against each other. Kephart and Conitzer \cite{Kephart2016} proposed that when reporting truthfully is costless and misreporting is costly, the revelation principle can fail to hold.

Different from these criticisms, this paper aims to propose another failure of the revelation principle. The paper is organized as follows: Section 2 analyses two formats of strategy and points out that the action format is very common in many practical cases. Section 3 proposes the main result, \emph{i.e.}, the revelation principle fails when each agent's strategy is action-format. Section 4 draws conclusions. Notations about mechanism design theory are given in Appendix, which are cited from Ref \cite{MWG1995}.

\section{Two formats of strategy}
\textbf{Note 1:} In any game, the format of each agent's strategy is either an informational message or a realistic action. $\Box$

Although the note looks naive, it is not trivial. The reason why we highlight the distinction of two formats of strategy is that the revelation principle does not hold for the case of action-format strategies, as will be discussed deeply in Section 3. For simplification, in the following discussions we simply assume that in any game all agents' strategies are of the same format, \emph{i.e.}, we omit the case in which some agents' strategies are message-format and other agents' strategies are action-format. Next, we will deeply investigate the two formats of strategy respectively.

\subsection{Case 1: Mechanism with message-format strategies}
\textbf{Definition 1:}  
A message-format strategy is represented by an informational message. The information itself contained in the message is just the strategy, which does not need to be carried out realistically in the mechanism.

Practically, only in some restricted cases can each agent's strategy be described as pure information and represented by an informational message. For example, let us consider a chess game, then each player's strategy is message-format, since it is a strategic plan about how to play chess. Similarly, the strategy in a war simulation game is also message-format, since it contains military \emph{plans} of players.

\textbf{Definition 2:} 
Given a social choice function $f$, suppose a mechanism $\Gamma=(S_{1}, \cdots, S_{I}, g(\cdot))$ implements it in Bayesian Nash equilibrium with message-format strategies. To clearly describe the message format, we denote each agent $i$'s strategy set $S_{i}$ as $M_{i}$, and each strategy function as $m_{i}(\cdot):\Theta_{i} \rightarrow M_{i}$, where $\Theta_{i}$ is agent $i$'s type set. The outcome function $g(\cdot)$ is denoted as $g_{m}(\cdot): M_{1}\times\cdots\times M_{I}\rightarrow X$, where the parameters are message-format strategies and $X$ is the set of outcomes. Hence, the mechanism $\Gamma$ is rewritten as $\Gamma_{m}=(M_{1}, \cdots, M_{I}, g_{m}(\cdot))$. The game induced by $\Gamma_{m}$ is denoted as $G_{m}$, which works in a one-stage manner:\\
\emph{Step 1:} By using the strategy function $m_{i}(\cdot):\Theta_{i} \rightarrow M_{i}$, each agent $i$ with private type $\theta_{i}$ sends a message $m_{i}(\theta_{i})$ to the designer. \footnote{In the following discussions, we denote each agent $i$'s true type as $\theta_{i}$, and his any possible type as $\hat{\theta}_{i}\in\Theta_{i}$.}\\
\emph{Step 2:} The mechanism $\Gamma_{m}$ yields the outcome $g_{m}(m_{1}(\theta_{1}), \cdots, m_{I}(\theta_{I}))$.\\
Here, each agent $i$'s utility is denoted as $u_{i}(g_{m}(m_{1}(\theta_{1}), \cdots, m_{I}(\theta_{I})), \theta_{i})$.

\textbf{Definition 3:} 
Suppose the game $G_{m}$ has a Bayesian Nash equilibrium, denoted as $m^{*}(\cdot)=(m^{*}_{1}(\cdot), \cdots, m^{*}_{I}(\cdot))$, \emph{i.e.}, for all $i$ and all $\theta_{i}\in\Theta_{i}$, $\hat{m}_{i}\in M_{i}$,
\begin{equation*}
  E_{\theta_{-i}}[u_{i}(g_{m}(m^{*}_{i}(\theta_{i}),m^{*}_{-i}(\theta_{-i})),\theta_{i})|\theta_{i}]
  \geq
  E_{\theta_{-i}}[u_{i}(g_{m}(\hat{m}_{i},m^{*}_{-i}(\theta_{-i})),\theta_{i})|\theta_{i}].
\end{equation*}
Consider this equilibrium, there is a  compound mapping $g_{m}(m^{*}(\cdot)):\Theta\rightarrow X$ from agents' possible types $\hat{\theta}=(\hat{\theta}_{1}, \cdots, \hat{\theta}_{I})\in\Theta$ into the outcome $g_{m}(m^{*}(\hat{\theta}))$, which is equal to $f(\hat{\theta})$ for any $\hat{\theta}\in\Theta$. Based on the compound mapping, we construct a direct mechanism $\bar{\Gamma}_{m}=(\Theta_{1}, \cdots, \Theta_{I}, g_{m}(m^{*}(\cdot)))$ with message-format strategies.
\footnote{Although $g_{m}(m^{*}(\hat{\theta}))=f(\hat{\theta})$ for any $\hat{\theta}\in\Theta$, the outcome function of the constructed direct mechanism $\bar{\Gamma}_{m}$ must be the compound mapping $g_{m}(m^{*}(\cdot)):\Theta\rightarrow X$ instead of $f(\cdot):\Theta\rightarrow X$. The reason is straightforward: if the outcome function of $\bar{\Gamma}_{m}$ is simply written as $f(\cdot)$, then $\bar{\Gamma}_{m}$ will become a naive direct mechanism $(\Theta_{1}, \cdots, \Theta_{I}, f(\cdot))$, which is irrelevant to the mechanism $\Gamma_{m}=(M_{1},\cdots,M_{I}, g_{m}(\cdot))$ which implements $f$ in equilibrium. Indeed, the formula $g_{m}(m^{*}(\hat{\theta}))=f(\hat{\theta})$ (for any $\hat{\theta}\in\Theta$) is the result of the mechanism $\Gamma_{m}$, and the naive direct mechanism $(\Theta_{1}, \cdots, \Theta_{I}, f(\cdot))$ cannot implement $f(\cdot)$ in Bayesian Nash equilibrium at all.}

\textbf{Definition 4:} 
The direct mechanism $\bar{\Gamma}_{m}$ induces a \emph{one-stage} direct game $\bar{G}_{m}$, which works as follows:\\
\emph{Step 1:} Each agent $i$ with private type $\theta_{i}$ individually reports an arbitrary type $\hat{\theta}_{i}\in\Theta_{i}$. Here, each agent $i$ does not need to be ``\emph{honest}'', \emph{i.e.}, $\hat{\theta}_{i}$ can be different from agent $i$'s private type $\theta_{i}$.\\
\emph{Step 2:} By using the equilibrium strategy functions $m^{*}(\cdot)=(m^{*}_{1}(\cdot), \cdots, m^{*}_{I}(\cdot))$, the direct mechanism $\bar{\Gamma}_{m}$ calculates $m^{*}(\hat{\theta})=(m^{*}_{1}(\hat{\theta}_{1}), \cdots, m^{*}_{I}(\hat{\theta}_{I}))$, and then yields the outcome $g_{m}(m^{*}(\hat{\theta}))$.

\textbf{Note 2:} Obviously, the calculated results $m^{*}(\hat{\theta})=(m^{*}_{1}(\hat{\theta}_{1}), \cdots, m^{*}_{I}(\hat{\theta}_{I}))$ are pure information, and hence are message-format. Actually, only when each agent $i$'s strategy $m_{i}(\cdot)$ is message-format can $m^{*}(\hat{\theta})$ be \emph{legal} parameters of the outcome function $g_{m}(\cdot)$.

\textbf{Note 3:} By Definition 3, $m^{*}(\cdot)=(m^{*}_{1}(\cdot), \cdots, m^{*}_{I}(\cdot))$ is the equilibrium of the game $G_{m}$, then $m^{*}_{i}(\theta_{i})$ is the optimal choice for each agent $i$ given that all other agents send $m^{*}_{-i}(\theta_{-i})$. Therefore, in the direct game $\bar{G}_{m}$, each agent $i$ will find truth-telling $\hat{\theta}_{i}=\theta_{i}$ to be the optimal choice given that the others agents tell the truth $\hat{\theta}_{-i}= \theta_{-i}$, and the final outcome will be $g_{m}(m^{*}(\theta))$, which is equal to $f(\theta)$ for all $\theta\in\Theta$. Thus, for the case of message-format strategies, truth-telling is a Bayesian Nash equilibrium of the direct game $\bar{G}_{m}$. Consequently, \emph{the revelation principle holds when each agent's strategy is message-format}. $\Box$

\subsection{Case 2: Mechanism with action-format strategies}
\textbf{Definition 5:} 
An action-format strategy is represented by a realistic action, which should be performed practically.

In many common cases each agent's strategy must be described as a realistic action instead of an informational message. For example, let us consider a tennis game, then each player's strategy is his realistic action of playing tennis, but not any imaginary plan about how to play tennis. Compared with the message-format strategy in a war simulation game, the strategy in a real war is action-format, since it contains military \emph{actions} of armies.


\textbf{Definition 6:} 
Given a social choice function $f$, suppose a mechanism $\Gamma=(S_{1}, \cdots, S_{I}, g(\cdot))$ implements it in Bayesian Nash equilibrium with action-format strategies.
To clearly describe the action format, we denote each agent $i$'s strategy set $S_{i}$ as $A_{i}$, and each strategy function as $a_{i}(\cdot):\Theta_{i}\rightarrow A_{i}$. The outcome function $g(\cdot)$ is denoted as $g_{a}(\cdot):A_{1}\times\cdots\times A_{I}\rightarrow X$, where the parameters are action-format strategies. Hence, the mechanism $\Gamma$ is rewritten as $\Gamma_{a}=(A_{1}, \cdots, A_{I}, g_{a}(\cdot))$. The game induced by $\Gamma_{a}$ is denoted as $G_{a}$, which works in a one-stage manner: \\
\emph{Step 1:} By using the action-format strategy function $a_{i}(\cdot):\Theta_{i}\rightarrow A_{i}$, each agent $i$ with private type $\theta_{i}$ performs the action-format strategy $a_{i}(\theta_{i})$. \footnote{For the case of action-format strategies, the designer \emph{observes} the performance of each agent's action. As a comparison, for the case of message-format strategies, the designer \emph{receives} each agent's message.}\\
\emph{Step 2:} The mechanism $\Gamma_{a}$ yields the outcome $g_{a}(a_{1}(\theta_{1}), \cdots, a_{I}(\theta_{I}))$.
\footnote{If in the mechanism $\Gamma_{a}$ some agent $i$ only declares a message about how to perform an action but does not realistically perform it, then this declaration is meaningless.}\\
Here, each agent $i$'s utility is denoted as $u_{i}(g_{a}(a_{1}(\theta_{1}), \cdots, a_{I}(\theta_{I})), \theta_{i})$.

\textbf{Definition 7:} 
Suppose the game $G_{a}$ has a Bayesian Nash equilibrium $a^{*}(\cdot)=(a^{*}_{1}(\cdot), \cdots$, $a^{*}_{I}(\cdot))$, \emph{i.e.}, for all $i$ and all $\theta_{i}\in\Theta_{i}$, $\hat{a}_{i}\in A_{i}$,
\begin{equation*}
  E_{\theta_{-i}}[u_{i}(g_{a}(a^{*}_{i}(\theta_{i}),a^{*}_{-i}(\theta_{-i})),\theta_{i})|\theta_{i}]
  \geq
  E_{\theta_{-i}}[u_{i}(g_{a}(\hat{a}_{i},a^{*}_{-i}(\theta_{-i})),\theta_{i})|\theta_{i}].
\end{equation*}
Consider this equilibrium, there is a compound mapping $g_{a}(a^{*}(\cdot)):\Theta\rightarrow X$ from agents' possible types $\hat{\theta}=(\hat{\theta}_{1}, \cdots, \hat{\theta}_{I})\in\Theta$ into the outcome $g_{a}(a^{*}(\hat{\theta}))$, which is equal to $f(\hat{\theta})$ for any $\hat{\theta}\in\Theta$. Based on the compound mapping, we construct a direct mechanism $\bar{\Gamma}_{a}=(\Theta_{1}, \cdots, \Theta_{I}$, $ g_{a}(a^{*}(\cdot)))$ with action-format strategies.

\textbf{Definition 8:} 
According to Myerson \cite{Myerson1982}, the direct mechanism $\bar{\Gamma}_{a}$ induces a \emph{multistage} direct game $\bar{G}_{a}$, which works as follows: \\
\emph{Step 1:} Each agent $i$ with private type $\theta_{i}$ individually reports an arbitrary type $\hat{\theta}_{i}\in\Theta_{i}$. Here each agent does not need to be ``\emph{honest}'', \emph{i.e.}, $\hat{\theta}_{i}$ can be different from $\theta_{i}$.\\
\emph{Step 2:} The direct mechanism $\bar{\Gamma}_{a}$ returns a suggestion to each agent $i$, here the suggestion is just a message-format description of action $a^{*}_{i}(\hat{\theta}_{i})\in A_{i}$. In order to emphasize the format of the suggestion is a message, we denote the suggestion as $a^{m}_{i}(\hat{\theta}_{i})$;\\
\emph{Step 3:} Each agent $i$ individually performs an action $\hat{a}_{i}\in A_{i}$. Here each agent $i$ does not need to be ``\emph{obedient}'', $i.e.$, $\hat{a}_{i}$ can be different from $a^{m}_{i}(\hat{\theta}_{i})$.\\
\emph{Step 4:} After observing all actions $\hat{a}_{1}, \cdots, \hat{a}_{I}$ have been performed, the direct mechanism $\bar{\Gamma}_{a}$ yields the outcome $g_{a}(\hat{a}_{1}, \cdots, \hat{a}_{I})$.\\
Here, each agent $i$'s utility is denoted as $u_{i}(g_{a}(\hat{a}_{1}, \cdots, \hat{a}_{I}), \theta_{i})$.


\textbf{Note 4:} Generally speaking, each agent's private type is his privacy and is valuable to him. Consider Step 1 in Definition 8, each agent $i$ reports an arbitrary type either honestly or dishonestly. Note that choosing to be honest or dishonest is each agent's independent and private choice, which cannot be controlled by the designer. \emph{From each agent's perspective, if the utility of truth-telling is not strictly larger than the utility of false-telling, then any reasonable agent will certainly prefer false-telling}, because false-telling always hides his privacy. $\Box$

\textbf{Note 5:} Consider Step 3 in Definition 8, each agent performs an action after receiving a suggestion, either obediently or disobediently. Note that choosing to be obedient or disobedient is each agent's independent and open choice, which cannot be controlled by the designer. Since the designer is not a dictator in the framework of mechanism design theory, she has no power to punish any disobedient agent. Obviously, each agent will choose the choice which yields the higher utility. $\Box$

\section{Main results}
Consider the multistage direct game $\bar{G}_{a}$ induced by the direct mechanism $\bar{\Gamma}_{a}$ described in Definition 8. According to Myerson \cite{Myerson1982}, the strategy ``\emph{honest and obedient}'' is the Bayesian Nash equilibrium of the game $\bar{G}_{a}$: \emph{i.e.}, each agent $i$ not only honestly discloses his private type in Step 1 ($i.e.$, $\hat{\theta}_{i}=\theta_{i}$), but also obeys the suggestion in Step 3 ($i.e.$, $\hat{a}_{i}=a^{m}_{i}(\theta_{i})$). However, in this section we will point out that Myerson's conclusion will not hold when each agent's strategy is of an action format.

\textbf{Proposition 1:} For any given social choice function $f(\cdot): \Theta\rightarrow X$, suppose there is a mechanism that implements it in Bayesian Nash equilibrium, and each agent's strategy is of an action format. Then $f$ will not be truthfully implementable, \emph{i.e.}, in the multistage direct game induced by the corresponding direct mechanism, ``\emph{honest and obedient}'' will no longer be the Bayesian Nash equilibrium.

\textbf{Proof:} Suppose a mechanism $\Gamma_{a}=(A_{1}, \cdots, A_{I}, g_{a}(\cdot))$ implements the social choice function $f(\cdot): \Theta\rightarrow X$ in Bayesian Nash equilibrium, and the format of each agent's strategy is an action. By Definition 6 it induces a one-stage game $G_{a}$, the equilibrium of which is denoted as $a^{*}(\cdot)=(a^{*}_{1}(\cdot), \cdots, a^{*}_{I}(\cdot))$, and the outcome $g_{a}(a^{*}(\theta))=f(\theta)$ for any $\theta\in\Theta$. By Definition 7 and 8, there is a corresponding direct mechanism $\bar{\Gamma}_{a}=(\Theta_{1}, \cdots, \Theta_{I}, g_{a}(a^{*}(\cdot)))$ and a multistage direct game $\bar{G}_{a}$. In the direct game $\bar{G}_{a}$, each agent $i$ with private type $\theta_{i}$ independently makes two decisions: in Step 1 he chooses to be ``\emph{honest}'' or ``\emph{dishonest}'', and in Step 3 he chooses to be ``\emph{obedient}'' or ``\emph{disobedient}''.

For any agent $i$, if all other agents $1, \cdots, i-1, i+1, \cdots, I$ choose to be ``\emph{honest}'' (\emph{i.e.}, $\hat{\theta}_{-i}=\theta_{-i}$), in the following discussions we investigate which choice agent $i$ should choose in Step 1 of $\bar{G}_{a}$, either ``\emph{honest}'' or ``\emph{dishonest}''.

\emph{Case 1:} Suppose agent $i$ chooses to be ``\emph{honest}'' in Step 1 of $\bar{G}_{a}$, \emph{i.e.}, $\hat{\theta}_{i}=\theta_{i}$. Then in Step 2 of $\bar{G}_{a}$, the suggestion to agent $i$ will be $a^{m}_{i}(\theta_{i})$, and the suggestions to the rest agents will be $a^{m}_{-i}(\theta_{-i})$. Since the Bayesian Nash equilibrium strategy of $G_{a}$ is $a^{*}(\cdot)=(a^{*}_{1}(\cdot), \cdots, a^{*}_{I}(\cdot))$, then in Step 3 of $\bar{G}_{a}$, choosing to be ``\emph{obedient}'' (\emph{i.e.}, $\hat{a}_{i}=a^{*}_{i}(\theta_{i})$) will be the optimal choice of agent $i$ if all other agents choose to be ``\emph{obedient}'' (\emph{i.e.}, $\hat{a}_{-i}=a^{*}_{-i}(\theta_{-i})$). In Step 4 of $\bar{G}_{a}$, the direct mechanism $\bar{\Gamma}_{a}$ will yield the outcome $g_{a}(a^{*}(\theta))$, which is equal to $f(\theta)$ for all $\theta\in\Theta$.

\emph{Case 2:} Suppose in Step 1 of $\bar{G}_{a}$, agent $i$ wants to hide his privacy and chooses to be ``\emph{dishonest}'', \emph{i.e.}, $\hat{\theta}_{i}\neq\theta_{i}$. Obviously, in Step 2 of $\bar{G}_{a}$, the suggestion to agent $i$ will be $a^{m}_{i}(\hat{\theta}_{i})\neq a^{m}_{i}(\theta_{i})$, and the suggestions to the others will still be $a^{m}_{-i}(\theta_{-i})$.
Therefore, if all other agents choose ``\emph{obedient}'' (\emph{i.e.}, performing $\hat{a}_{-i}=a^{*}_{-i}(\theta_{-i})$), then the optimal choice of agent $i$ in Step 3 of $\bar{G}_{a}$ should be ``\emph{disobedient}'' (\emph{i.e.}, not obeying the suggestion $a^{m}_{i}(\hat{\theta}_{i})$ but still performing $\hat{a}_{i}=a^{*}_{i}(\theta_{i})$).
Although the designer can find the agent $i$ is disobedient in Step 4 of $\bar{G}_{a}$, she cannot punish him according to Note 5. Consequently, in Step 4 of $\bar{G}_{a}$, the direct mechanism $\bar{\Gamma}_{a}$ will still yield the outcome $g_{a}(a^{*}(\theta))$, which is equal to $f(\theta)$ for all $\theta\in\Theta$.

It should be emphasized that although in Case 2 the agent $i$ chooses to be ``\emph{disobedient}'' in Step 3 of $\bar{G}_{a}$, what he really performs is still consistent with his private type, \emph{i.e.}, $\hat{a}_{i}=a^{*}_{i}(\theta_{i})$, the same as what agent $i$ performs in Case 1. As a result, the social choice function $f(\cdot)$ is still implemented in Bayesian Nash equilibrium by the constructed direct mechanism $\bar{\Gamma}_{a}=(\Theta_{1}, \cdots, \Theta_{I}, g_{a}(a^{*}(\cdot)))$.

From agent $i$'s perspective, his privacy is disclosed in Case 1 and hidden in Case 2. Thus, Case 2 will be strictly better than Case 1, because agent $i$ can obtain the same outcome $g_{a}(a^{*}(\theta))$ in the two cases but protects his privacy in Case 2. Consequently, ``\emph{honest and obedient}'' will no longer be the Bayesian Nash equilibrium of the direct mechanism. Furthermore, since Case 2 holds for any agent $i$, it can be generalized to every agent as follows.

\emph{Case 3:} Suppose each agent $i$ with private type $\theta_{i}$ chooses to be ``\emph{dishonest}'' in Step 1 of $\bar{G}_{a}$ (\emph{i.e.} reporting a false type $\hat{\theta}_{i}\neq\theta_{i}$), and then chooses to be ``\emph{disobedient}'' in Step 3 of $\bar{G}_{a}$, \emph{i.e.}, not obeying the designer's suggestion $a^{m}_{i}(\hat{\theta}_{i})$ but still performing the action $\hat{a}_{i}=a^{*}_{i}(\theta_{i})$ which is consistent with his private type. Since the Bayesian Nash equilibrium strategy of $G_{a}$ is $a^{*}(\cdot)=(a^{*}_{1}(\cdot), \cdots, a^{*}_{I}(\cdot))$, then the final outcome of $\bar{G}_{a}$ will still be $g_{a}(a^{*}(\theta))$, which is equal to $f(\theta)$ for all $\theta\in\Theta$. Put differently, $f(\cdot)$ is implemented by the constructed direct mechanism $\bar{\Gamma}_{a}=(\Theta_{1}, \cdots, \Theta_{I}, g_{a}(a^{*}(\cdot)))$ in Bayesian Nash equilibrium, \emph{not truthfully, but dishonestly and disobediently}.

Compared with Case 1, in Case 3 each agent obtains the same outcome $g_{a}(a^{*}(\theta))$ and at the same time protects his privacy. Therefore, ``\emph{dishonest and disobedient}'' will be strictly better than ``\emph{honest and obedient}'' from each agent's perspective.

To sum up, when the format of each agent's strategy is an action, ``\emph{dishonest and disobedient}'' instead of ``\emph{honest and obedient}'' will be the Bayesian Nash equilibrium of the direct mechanism $\bar{\Gamma}_{a}=(\Theta_{1}, \cdots, \Theta_{I}, g_{a}(a^{*}(\cdot)))$, which means the revelation principle does not hold. $\Box$

\section{Conclusions}
In this paper, we propose that in any game there are two formats of strategy (\emph{i.e.}, an informational message or a realistic action), and the action format is very common in many practical cases. In Section 2.1 we point out that the revelation principle holds when the format of each agent's strategy is a message. However, when the format of each agent's strategy is an action, ``\emph{honest and obedient}'' will no longer be the Bayesian Nash equilibrium of the direct mechanism, actually the social choice function can only be implemented ``\emph{dishonestly and disobediently}'' in Bayesian Nash equilibrium by the direct mechanism. Therefore, the revelation principle fails when each agent's strategy is of action format.

\section*{Appendix: Notations and Definitions \cite{MWG1995}}
Let us consider a setting with one designer and $I$ agents indexed by $i=1,\cdots,I$. Each agent $i$ privately observes his \emph{type} $\theta_{i}$ that determines his preference over elements in an outcome set $X$. The set of possible types for agent $i$ is denoted as $\Theta_{i}$. The vector of agents' types $\theta=(\theta_{1},\cdots,\theta_{I})$ is drawn from set $\Theta=\Theta_{1}\times\cdots\times\Theta_{I}$ according to probability density $\phi(\cdot)$, and each agent $i$'s \emph{utility function} over the outcome $x\in X$ given his type $\theta_{i}$ is $u_{i}(x,\theta_{i})\in\mathbb{R}$.

\textbf{Definition 23.B.1} A \emph{social choice function} (SCF) is a function $f:\Theta_{1}\times\cdots\times\Theta_{I}\rightarrow X$
that, for each possible profile of the agents' types $\theta_{1},\cdots,\theta_{I}$, assigns a collective choice $f(\theta_{1},\cdots,\theta_{I})\in X$.

\textbf{Definition 23.B.3} A \emph{mechanism} $\Gamma=(S_{1},\cdots,S_{I},g(\cdot))$ is a collection of $I$ strategy sets $S_{1},\cdots,S_{I}$ and an outcome function $g:S_{1}\times\cdots\times S_{I}\rightarrow X$.

The mechanism combined with possible types $(\Theta_{1},\cdots,\Theta_{I})$, the probability density $\phi(\cdot)$ over the possible realizations of $\theta\in\Theta_{1}\times\cdots\times\Theta_{I}$, and utility functions $(u_{1}, \cdots, u_{I})$ defines a Bayesian game of incomplete information. The strategy function of each agent $i$ in the game induced by $\Gamma$ is a private function $s_{i}(\theta_{i}): \Theta_{i}\rightarrow S_{i}$. Each strategy set $S_{i}$ contains agent $i$'s possible strategies. The outcome function $g(\cdot)$ describes the rule for how agents' strategies get turned into a social choice.

\textbf{Definition 23.B.5} A \emph{direct mechanism} is a mechanism $\bar{\Gamma}=(\bar{S}_{1}, \cdots, \bar{S}_{I}$, $\bar{g}(\cdot))$ in which $\bar{S}_{i}=\Theta_{i}$ for all $i$ and $\bar{g}(\theta)=f(\theta)$ for all $\theta\in\Theta$. \footnote{The bar symbol is used to distinguish the direct mechanism from the indirect mechanism.}

\textbf{Definition 23.D.1} A strategy profile $s^{*}(\cdot)=(s^{*}_{1}(\cdot),\cdots,s^{*}_{I}(\cdot))$ is a \emph{Bayesian Nash equilibrium} of mechanism $\Gamma=(S_{1},\cdots,S_{I},g(\cdot))$ if, for all $i$ and all $\theta_{i}\in\Theta_{i}$, $\hat{s}_{i}\in S_{i}$, there exists
\begin{equation*}
  E_{\theta_{-i}}[u_{i}(g(s^{*}_{i}(\theta_{i}),s^{*}_{-i}(\theta_{-i})),\theta_{i})|\theta_{i}]
  \geq
  E_{\theta_{-i}}[u_{i}(g(\hat{s}_{i},s^{*}_{-i}(\theta_{-i})),\theta_{i})|\theta_{i}].
\end{equation*}

\textbf{Definition 23.D.2} The mechanism
$\Gamma=(S_{1},\cdots,S_{I},g(\cdot))$ \emph{implements the social choice
function} $f(\cdot)$ \emph{in Bayesian Nash equilibrium} if there is a
Bayesian Nash equilibrium of $\Gamma$,
$s^{*}(\cdot)=(s^{*}_{1}(\cdot),\cdots,s^{*}_{I}(\cdot))$, such that
$g(s^{*}(\theta))=f(\theta)$ for all $\theta\in\Theta$.

\textbf{Definition 23.D.3} The social choice function $f(\cdot)$ is
\emph{truthfully implementable in Bayesian Nash equilibrium} (or \emph{Bayesian incentive compatible}) if
$\bar{s}^{*}_{i}(\theta_{i})=\theta_{i}$ (for all
$\theta_{i}\in\Theta_{i}$ and $i=1,\cdots,I$) is a Bayesian Nash
equilibrium of the direct mechanism $\Bar{\Gamma}=(\Theta_{1},\cdots,\Theta_{I}, f(\cdot))$. That is, if for
all $i=1,\cdots,I$ and all $\theta_{i}\in\Theta_{i}$, $\hat{\theta}_{i}\in \Theta_{i}$,
\begin{equation*}
  E_{\theta_{-i}}[u_{i}(f(\theta_{i},\theta_{-i}),\theta_{i})|\theta_{i}]
  \geq
  E_{\theta_{-i}}[u_{i}(f(\hat{\theta}_{i},\theta_{-i}),\theta_{i})|\theta_{i}].
\end{equation*}

\textbf{Proposition 23.D.1}: (\emph{The Revelation Principle for Bayesian Nash Equilibrium}) Suppose that there exists a mechanism $\Gamma=(S_{1},\cdots,S_{I},g(\cdot))$ that implements the social choice function $f(\cdot)$ in Bayesian Nash equilibrium. Then $f(\cdot)$ is truthfully implementable in Bayesian Nash equilibrium.


\end{document}